# Synthesis of multi-wavelength temporal phase-shifting algorithms optimized for high signal-to-noise ratio and high detuning robustness using the frequency transfer function

**Manuel Servin**[*]**, Moises Padilla and Guillermo Garnica**

*Centro de Investigaciones en Optica A. C., Loma del Bosque 115, Col. Lomas del Campestre, 37150 Leon Guanajuato, Mexico.*
[*]*mservin@cio.mx*

**Abstract:** Synthesis of single-wavelength temporal phase-shifting algorithms (PSA) for interferometry is well-known and firmly based on the frequency transfer function (FTF) paradigm. Here we extend the single-wavelength FTF-theory to dual and multi-wavelength PSA-synthesis when several simultaneous laser-colors are present. The FTF-based synthesis for dual-wavelength PSA (DW-PSA) is optimized for high signal-to-noise ratio and minimum number of temporal phase-shifted interferograms. The DW-PSA synthesis herein presented may be used for interferometric contouring of discontinuous industrial objects. Also DW-PSA may be useful for DW shop-testing of deep free-form aspheres. As shown here, using the FTF-based synthesis one may easily find explicit DW-PSA formulae optimized for high signal-to-noise and high detuning robustness. To this date, no general synthesis and analysis for temporal DW-PSAs has been given; only *had-hoc* DW-PSAs formulas have been reported. Consequently, no explicit formulae for their spectra, their signal-to-noise, their detuning and harmonic robustness has been given. Here for the first time a fully general procedure for designing DW-PSAs (or triple-wavelengths PSAs) with desire spectrum, signal-to-noise ratio and detuning robustness is given. We finally generalize DW-PSA to higher number of wavelength temporal PSAs.

09-March-2016 Centro de Investigaciones en Optica, Mexico.

## 1. Introduction

As far as we know, the first researcher to use dual-wavelength (DW) interferometry was Wyant in 1971 [1]. Wyant used two fixed laser-wavelengths $\lambda_1$ and $\lambda_2$ to test an optical surface with an equivalent wavelength of $\lambda_{eq} = \lambda_1\lambda_2 / |\lambda_1 - \lambda_2|$ [2]. Thus typically $\lambda_{eq}$ is much larger than either $\lambda_1$ or $\lambda_2$ ($\lambda_{eq} >> \{\lambda_1, \lambda_2\}$). Double-wavelength (DW) interferometry was improved by Polhemus [3] and Cheng [4,5] using digital temporal phase-shifting.

On the other hand, Onodera et al. [6] used spatial-carrier, double-wavelength digital-holography (DW-DH) and Fourier interferometry for phase-demodulation. This in turn was followed by a large number of multi-wavelength digital-holographic (DH) Fourier phase-demodulation methods in such diverse applications as interferometric contouring [7], phase-imaging [8], chromatic aberration compensation in microscopy [9]; single hologram DW microscopy [10]; comb multi-wavelength laser for extended range optical metrology [11], and a two-steps digital-holography for image quality improvement [12].

More recently temporal dual-wavelength phase-shifting algorithms (DW-PSAs) have been reworked by Abdelsalam et al. [14]. Even though Abdelsalam et al. give working PSA formulas they do not estimate their spectra, their signal-to noise ratio, or their detuning and harmonics robustness. Kumar [15] and Baranda [16] also provided valid temporal PSA formulas but also failed to characterize their PSAs in terms of signal-to-noise, detuning and harmonic rejection. Another different approach was followed by Kulkarni and Rastogi [16] in which they have demodulated the two interesting phases by fitting a low-order polynomial to each phase. Their approach [17] worked well for the example provided but we think their method could easily cross-talk between fitted polynomials for complicated modulating phases [17]. Yet another approach by Zhang et al. was published [18-19]. Zhang used a simultaneous two-steps [18], and principal component interferometry [19] to solve the dual-wavelength phase-shifting measurement. Zhang et al. used 32 randomly phase-shifted interferograms [19]. Even though Zhang [19] could demodulate the two phases, they used 32 phase-shifted temporal interferograms. All these works on temporal DW-PSA [2-5,14-19] have given just specific DW-PSAs without explicit formulae for their spectra, signal-to-noise, detuning and harmonic robustness.

In contrast to previous *ad-hoc* temporal DW-PSA formulas, here we give a general theory for synthesizing DW-PSAs formalizing their spectrum, their signal-to-noise, and their detuning-harmonic robustness. At the risk of being repetitive, we emphasize that we are not just giving particular DW-PSAs formulas as previously done [2-5, 14-19]. Here we are giving a general FTF-theory for synthesizing DW-PSA giving with explicit formulae for the most important characteristics of any PSA: spectra, signal-to-noise, detuning and harmonic robustness.

## 2. Spatial-carrier phase-demodulation for Dual-wavelength (DW) interferometry

Dual-wavelength digital-holography (DW-DH) is well understood and widely used [6-10]. As shown in Fig. 1, in DW-DH the two lasers beams are tilted to introduce spatial-carrier fringes [7]. In Fig. 1 both lasers beams are tilted in the $x$ direction, but in general, for a better use of the Fourier space, one may tilt them independently along the $x$ and $y$ directions [11-14].

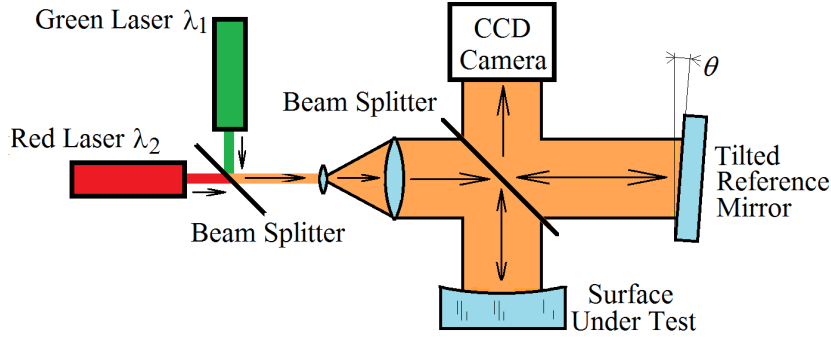

Fig.1 Schematics for DW-DH with a single tilted reference mirror [6]. The orange-light corresponds to the spatial superposition of the red and green lasers.

The DW-DH obtained at the CCD camera in Fig.1 may be modeled by,

$$I(x,y,t) = a(x,y) + b_1(x,y)\cos[\varphi_1(x,y) + u_1 x] + b_2(x,y)\cos[\varphi_2(x,y) + u_2 x]. \qquad (1)$$

Here $u_1 x = x(2\pi/\lambda_1)\tan(\theta)$ and $u_2 x = x(2\pi/\lambda_2)\tan(\theta)$ are the spatial-carriers of the DW-DH. The reference mirror angle along the $x$ axis is $\theta$. The searched phases are $\varphi_1(x,y) = (2\pi/\lambda_1)W_1(x,y)$ and $\varphi_2(x,y) = (2\pi/\lambda_2)W_2(x,y)$; being $W_1(x,y)$ and $W_2(x,y)$ the measuring wavefronts. Figure 2 shows a Fourier spectrum of Eq. (1).

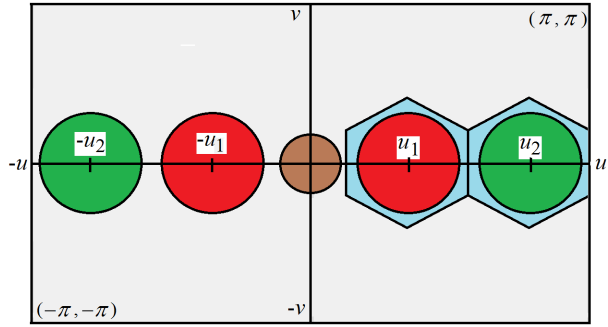

Fig. 2. The hexagons represent the spatial filters, which demodulate the phases $\varphi_1$ and $\varphi_2$.

The two hexagons in Fig. 2 are the quadrature filters that passband the desired analytic signals. After filtering, the inverse Fourier transform find the demodulated phases [1]. The advantage of DW-DH is that only one digital-hologram is needed to obtain $\{\varphi_1, \varphi_2\}$; however its drawback is that just a fraction of the Fourier space $(u,v) \in [-\pi,\pi] \times [-\pi,\pi]$ is used (Fig 2). This limitation makes DW-DH not suitable for measuring discontinuous industrial objects [7]. In contrast, in DW-PSAs the full Fourier spectrum $(u,v) \in [-\pi,\pi] \times [-\pi,\pi]$ may be used.

## 3. Dual-wavelength (DW) temporal-carrier phase-shifting interferometry

The temporal phase-shifting double-wave interferogram may be modeled as,

$$I(x,y,t) = a(x,y) + b_1(x,y)\cos\left[\varphi_1(x,y) + \left(\frac{2\pi}{\lambda_1}d\right)t\right] + b_2(x,y)\cos\left[\varphi_2(x,y) + \left(\frac{2\pi}{\lambda_2}d\right)t\right]. \quad (2)$$

Where $t \in (-\infty, \infty)$, and $\varphi_1(x,y) = (2\pi/\lambda_1)W_1(x,y)$, $\varphi_2(x,y) = (2\pi/\lambda_2)W_2(x,y)$ are the measuring phases. The parameter $d$ is the PZT-step. The fringes background is $a(x,y)$ and the lasers power must be about the same to obtain high fringe contrast: $b_1(x,y) \approx b_2(x,y)$. Figure 3 shows one possible set-up for a DW temporal phase-shifting interferometer.

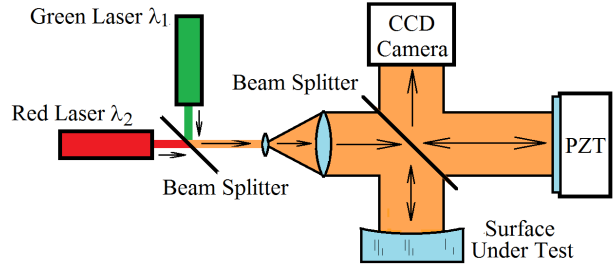

Fig. 3. A schematic example of a DW temporal-carrier interferometer [2-5] for surface measured with equivalent wavelength $\lambda_{eq}$; the piezoelectric transducer is PZT.

The motivation of using 2-wavelengths $\lambda_1$ and $\lambda_2$ (in spatial or temporal interferometry) is that interferometric measurements can be made with an equivalent wavelength $\lambda_{eq}$ [2-19],

$$\lambda_{eq} = \frac{\lambda_1 \lambda_2}{|\lambda_1 - \lambda_2|}; \quad \lambda_{eq} \gg (\lambda_1 \text{ or } \lambda_2). \quad (3)$$

With large $\lambda_{eq}$ one may measure deeper surface discontinuities or topographies than using either $\lambda_1$ or $\lambda_2$ alone [2-19]. For a given PZT-step $d$, the two angular-frequencies (in radians per interferogram) are given by,

$$\omega_1 = \frac{2\pi}{\lambda_1}d, \quad \text{and} \quad \omega_2 = \frac{2\pi}{\lambda_2}d. \quad (4)$$

Using this equation one may rewrite Eq. (2) as,

$$I(x,y,t) = a(x,y) + b_1(x,y)\cos[\varphi_1(x,y) + \omega_1 t] + b_2(x,y)\cos[\varphi_2(x,y) + \omega_2 t], \quad (5)$$

Here we have 5 unknowns: namely, $\{a, b_1, b_2, \varphi_1, \varphi_2\}$. Therefore we need at least 5 phase-shifted interferograms to obtain a solution for $\varphi_1(x,y)$ and $\varphi_2(x,y)$. These are given by:

$$\begin{aligned}
I_0(x,y) &= a + b_1\cos[\varphi_1] + b_2\cos[\varphi_2], \\
I_1(x,y) &= a + b_1\cos[\varphi_1 + \omega_1] + b_2\cos[\varphi_2 + \omega_2], \\
I_2(x,y) &= a + b_1\cos[\varphi_1 + 2\omega_1] + b_2\cos[\varphi_2 + 2\omega_2], \\
I_3(x,y) &= a + b_1\cos[\varphi_1 + 3\omega_1] + b_2\cos[\varphi_2 + 3\omega_2], \\
I_4(x,y) &= a + b_1\cos[\varphi_1 + 4\omega_1] + b_2\cos[\varphi_2 + 4\omega_2].
\end{aligned} \quad (6)$$

For clarity, most $(x,y)$ coordinates were omitted.

## 4. Fourier-spectrum for temporal DW-PSAs

The Fourier transform of the temporal interferogram (with $t \in (-\infty, \infty)$) in Eq. (5) is:

$$I(\omega) = a\,\delta(\omega) + \frac{b_1}{2}\left[e^{i\varphi_1}\delta(\omega-\omega_1) + e^{-i\varphi_1}\delta(\omega+\omega_1)\right] + \frac{b_2}{2}\left[e^{i\varphi_2}\delta(\omega-\omega_2) + e^{-i\varphi_2}\delta(\omega+\omega_2)\right]. \quad (7)$$

All $(x,y)$ were omitted. As mentioned, $\omega_1 = (2\pi/\lambda_1)d$ and $\omega_2 = (2\pi/\lambda_2)d$ are the two temporal-carrier frequencies in radians/interferogram; Fig. 4 shows this spectrum.

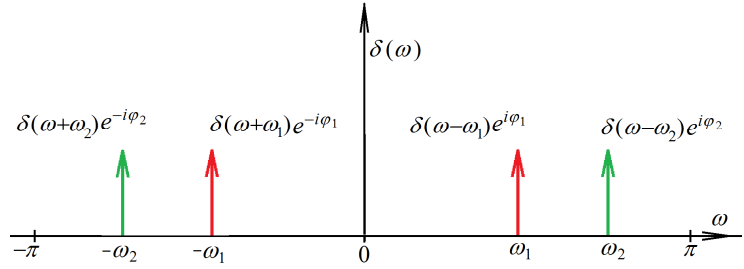

Fig. 4. Fourier spectrum of the DW temporal-carrier interferograms.

Figure 5 shows two ideal quadrature filters $H_1(\omega)$ and $H_2(\omega)$ that could passband the desired analytic signals $\delta(\omega-\omega_1)\exp(i\,\varphi_1)$ and $\delta(\omega-\omega_2)\exp(i\,\varphi_2)$. Note how each filter is able to passband the desired signals from the same temporal interferograms.

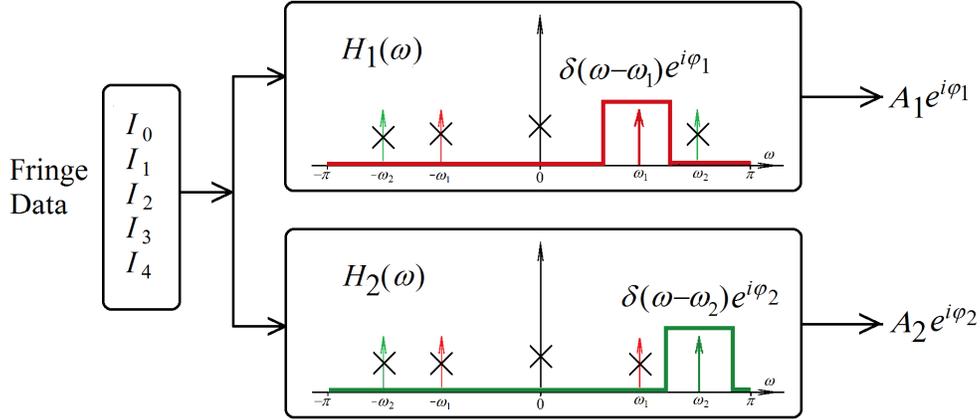

Fig. 5. Ideal spectra of two filters that passband the desired signals $\exp(i\,\varphi_1)$ and $\exp(i\,\varphi_2)$ from temporal phase-shifted interferograms; all crossed Dirac deltas are filtered-out.

## 5. Synthesis of DW-PSAs using 5-step temporal interferograms

The rectangular filters in Fig. 5 require a large number of temporal interferograms [1]. However, using the FTF we can synthesize 5-step bandpass filters by allocating 4 spectral-zeroes at frequencies $\{-\omega_2, -\omega_1, 0, \omega_2\}$ for $H_1(\omega)$, and at $\{-\omega_2, -\omega_1, 0, \omega_1\}$ for $H_2(\omega)$ as,

$$\begin{aligned}H_1(\omega) &= \left(1-e^{i\omega}\right)\left[1-e^{i(\omega+\omega_2)}\right]\left[1-e^{i(\omega-\omega_2)}\right]\left[1-e^{i(\omega+\omega_1)}\right], \\ H_2(\omega) &= \left(1-e^{i\omega}\right)\left[1-e^{i(\omega-\omega_1)}\right]\left[1-e^{i(\omega+\omega_1)}\right]\left[1-e^{i(\omega+\omega_2)}\right].\end{aligned} \quad (8)$$

From Eq. (8) one sees that by design, $I(\omega)H_1(\omega)$ passband the signal $\exp(i\varphi_1)\delta(\omega-\omega_1)$, while $I(\omega)H_2(\omega)$ bandpass $\exp(i\varphi_2)\delta(\omega-\omega_2)$. The impulse responses $h_1(t)$ and $h_2(t)$ of these two quadrature filters are then given by:

$$h_1(t) = F^{-1}\{H_1(\omega)\} = \sum_{n=0}^{4} c1_n(\omega_1,\omega_2)\,\delta(t-n),$$
$$h_2(t) = F^{-1}\{H_2(\omega)\} = \sum_{n=0}^{4} c2_n(\omega_1,\omega_2)\,\delta(t-n). \quad (9)$$

Here $c1_n(\omega_1,\omega_2)$ and $c2_n(\omega_1,\omega_2)$ are complex-valued coefficients that depend on the frequencies $\{\omega_1,\omega_2\}$. Having $h_1(t)$ and $h_2(t)$, we obtain the two searched DW-PSAs as,

$$\frac{1}{2}H_1(\omega_1)b(x,y)e^{i\varphi_1(x,y)} = \sum_{n=0}^{4} c1_n(\omega_1,\omega_2)\,I_n(x,y),$$
$$\frac{1}{2}H_2(\omega_2)b(x,y)e^{i\varphi_2(x,y)} = \sum_{n=0}^{4} c2_n(\omega_1,\omega_2)\,I_n(x,y). \quad (10)$$

The explicit 5-step DW-PSA formula to estimate $\varphi_1(x,y)$ is:,

$$A_1\exp(i\varphi_1) = -e^{i\omega_2}I_0 + c1_1(\omega_1,\omega_2)I_1 - c1_2(\omega_1,\omega_2)I_2 + c1_3(\omega_1,\omega_2)I_3 - e^{i(\omega_2-\omega_1)}I_4,$$
$$c1_1(\omega_1,\omega_2) = 1 + e^{i\omega_2} + e^{2i\omega_2} + e^{i(\omega_2-\omega_1)},$$
$$c1_2(\omega_1,\omega_2) = 1 + e^{i\omega_2} + e^{2i\omega_2} + e^{i(\omega_2-\omega_1)} + e^{-i\omega_1} + e^{i(2\omega_2-\omega_1)}, \quad (11)$$
$$c1_3(\omega_1,\omega_2) = \left[1 + e^{-i\omega_1} + e^{-i(\omega_2+\omega_1)} + e^{i(\omega_2-\omega_1)}\right]e^{i\omega_2}.$$

Being $A_1 = (1/2)H_1(\omega_1)b(x,y)$. Conversely the 5-step DW-PSA to estimate $\varphi_2(x,y)$ is:

$$A_2\exp(i\varphi_2) = -e^{i\omega_1}I_0 + c2_1(\omega_1,\omega_2)I_1 - c2_2(\omega_1,\omega_2)I_2 + c2_3(\omega_1,\omega_2)I_3 - e^{i(\omega_1-\omega_2)}I_4,$$
$$c2_1(\omega_1,\omega_2) = 1 + e^{i\omega_1} + e^{2i\omega_1} + e^{i(\omega_1-\omega_2)},$$
$$c2_2(\omega_1,\omega_2) = 1 + e^{i\omega_1} + e^{2i\omega_1} + e^{i(\omega_1-\omega_2)} + e^{-i\omega_2} + e^{i(2\omega_1-\omega_2)}, \quad (12)$$
$$c2_3(\omega_1,\omega_2) = \left[1 + e^{-i\omega_2} + e^{-i(\omega_1+\omega_2)} + e^{i(\omega_1-\omega_2)}\right]e^{i\omega_1}.$$

Being $A_2 = (1/2)H_2(\omega_2)b(x,y)$. This is the basics for synthesizing DW-PSAs grounded on the FTF paradigm [1]. Previous papers on DW-PSAs [2-5,14-19] stop much shorter than this. They just show particular pairs of DW-PSAs [2-5,14-19] that work for just particular carriers, i.e. $(\omega_1,\omega_2) = (1.2, 2.9)$. In this section, we offered DW-PSAs (Eqs. (11)-(12)) which work well (find $\varphi1$ and $\varphi2$) for infinitely-continuous frequency-pairs $(\omega_1,\omega_2) \in [-\pi,\pi] \times [-\pi,\pi]$. Even if the theory of this paper would stop right here, this paper would contain a substantial improvement against current *ad-hoc* art in DW-PSA [2-5,14-19].

## 6. Signal-to-noise power-ratios for $H_1(\omega)$ and $H_2(\omega)$

Here we review the signal-to-noise power-ratio formulas for PSA quadrature filters [1]. The signal-to-noise power-ratios for $H_1(\omega)$ and $H_2(\omega)$ are given by [1]:

$$SNR_1 = \frac{|H_1(\omega_1)|^2}{\frac{1}{2\pi}\int_{-\pi}^{\pi}|H_1(\omega)|^2\,d\omega}, \quad SNR_2 = \frac{|H_2(\omega_2)|^2}{\frac{1}{2\pi}\int_{-\pi}^{\pi}|H_2(\omega)|^2\,d\omega}. \quad (13)$$

These 2 *SNR*-formulas give the power of the demodulated signals $|H_1(\omega_1)|^2$ and $|H_2(\omega_2)|^2$ divided by their total noise-power $(1/2\pi)\int|H_1(\omega)|^2 d\omega$ and $(1/2\pi)\int|H_2(\omega)|^2 d\omega$.

## 7. Non-optimized DW-PSA design for wavelengths $\lambda_1 = 632.8\,\text{nm}$ and $\lambda_2 = 532.0\,\text{nm}$

Let us assume that we use a typical temporal frequency of $\omega_1 = 2\pi/5$ radians per sample for the algorithm $H_1(\omega_1)e^{i\varphi_1(x,y)}$. Having made this choice for $\omega_1$, the frequency $\omega_2$ is set to

$$d = \omega_1\left(\frac{\lambda_1}{2\pi}\right) = \omega_2\left(\frac{\lambda_2}{2\pi}\right) \;\Rightarrow\; \omega_2 = \omega_1\left(\frac{\lambda_1}{\lambda_2}\right) \;\therefore\; \omega_2 = 1.49\,\frac{\text{radians}}{\text{sample}}, \quad (14)$$

and the required PZT-step is $d = 126.6\,\text{nm}$. The DW-FTFs for these two frequencies are:

$$\begin{aligned}
H_1(\omega) &= (1-e^{i\omega})\left[1-e^{i[\omega+1.49]}\right]\left[1-e^{i[\omega-1.49]}\right]\left[1-e^{i(\omega+1.26)}\right], \\
H_2(\omega) &= (1-e^{i\omega})\left[1-e^{i(\omega-1.26)}\right]\left[1-e^{i(\omega+1.26)}\right]\left[1-e^{i[\omega+1.49]}\right].
\end{aligned} \quad (15)$$

Figure 6 shows the magnitude plot of these two quadrature filters $H_1(\omega)$ and $H_2(\omega)$.

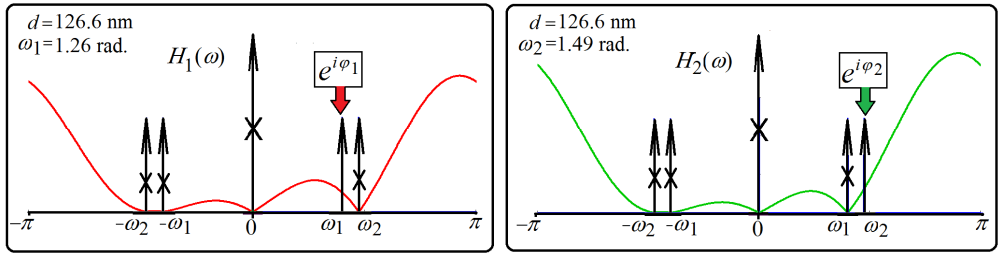

Fig. 6 Spectral plots for the two DW-PSA. The crossed Dirac deltas are the filter-out signals.
These two FTFs can demodulate $\varphi_1$ and $\varphi_2$ but with poor signal-to-noise performance.

The signal-to-noise [1] for the searched signals $H_1(\omega_1)\exp(i\varphi_1)$ and $H_2(\omega_2)\exp(i\varphi_2)$ are,

$$\frac{|H_1(\omega_1)|^2}{\frac{1}{2\pi}\int_{-\pi}^{\pi}|H_1(\omega)|^2 d\omega} = 0.94\,;\quad \frac{|H_2(\omega_2)|^2}{\frac{1}{2\pi}\int_{-\pi}^{\pi}|H_2(\omega)|^2 d\omega} = 1.04\,;\quad \omega_1=1.26;\,\omega_2=1.49. \quad (16)$$

For comparison, a 5-step least-squares PSA has a signal-to-noise power-ratio of 5 [1]. Thus $\omega_1 = 2\pi/5$ and $\omega_2 = 1.49$ were a bad choice; we can estimate $\varphi_1(x,y)$ and $\varphi_2(x,y)$ from the DW-PSAs in Eq. (11)-(12) using these temporal frequencies, but they are going to be noisy. Previous DW-PSAs efforts [2-5,14-19] only provide numeric-specific DW-PSAs formulas to obtain $\varphi_1(x,y)$ and $\varphi_2(x,y)$. However, they were silent about their Fourier spectra; their signal-to-noise; their harmonics rejection and their detuning robustness. All this useful and practical formulae are given here for the first time in terms of the FTF for designing DW-PSAs. Moreover, in contrast to previous art in DW-PSAs, Eq. (11) and Eq. (12) give infinite DW-PSA formulas for continuous pairs of temporal frequencies $(\omega_1,\omega_2)\in[-\pi,\pi]\times[-\pi,\pi]$.

## 8. Optimized joint signal-to-noise ratio synthesis for DW-PSAs

To find a better selection for $\omega_1 = (2\pi/\lambda_1)d$ and $\omega_2 = (2\pi/\lambda_2)d$, we construct a joint signal-to-noise ratio as,

$$G_{S/N}(d) = \left( \frac{|H_1(\omega_1)|^2}{\frac{1}{2\pi}\int_{-\pi}^{\pi}|H_1(\omega)|^2 d\omega} \right)\left( \frac{|H_2(\omega_2)|^2}{\frac{1}{2\pi}\int_{-\pi}^{\pi}|H_2(\omega)|^2 d\omega} \right); \quad d \in [0, \lambda_{eq}]. \qquad (17)$$

The function $G_{S/N}(d)$ is complicated and has many local maxima but, fortunately, it is one-dimensional. Thus we plot $G_{S/N}(d)$ and look for a good maximum, and take the PZT-step $d$. This PZT-step $d$ is used to find the two specific DW-PSA (Eqs. (11)-(12)) which solves the dual-wavelength interferometric problem.

## 9. Example of optimized DW-PSA synthesis for $\lambda_1 = 632.8\,\text{nm}$ and $\lambda_2 = 532\,\text{nm}$

The graph for the joint signal-to-noise ratio $G_{S/N}(d)$ with $\omega_1 = (2\pi/\lambda_1)d$, $\omega_2 = (2\pi/\lambda_2)d$ and $d \in [0, \lambda eq]$ is shown next.

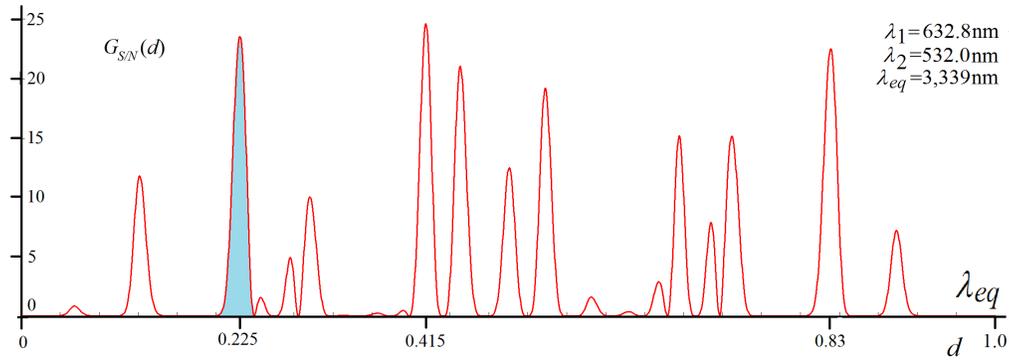

Fig. 7. Graph of $G_{S/N}(d)$. We kept the third local maximum at $d = 0.225\lambda_{eq} = 751\,\text{nm}$, for which $G_{S/N}(d) = 23.5$. Each DW-PSA thus have a signal-to-noise of $\sqrt{23.5} \approx 4.84$.

The first good-enough maximum is $G_{S/N}(0.225\lambda_{eq}) \approx 23.5$ (in blue), being $d = 0.225\lambda_{eq}$ or $d = 751\,\text{nm}$. Note that most of this graph is less than 20; i.e. $G_{S/N}(d) < 20$. This means that taking a PZT-step within $d \in [0, \lambda_{eq}]$ *at random*, the probability of landing in a very low signal-to-noise point is *very high*.

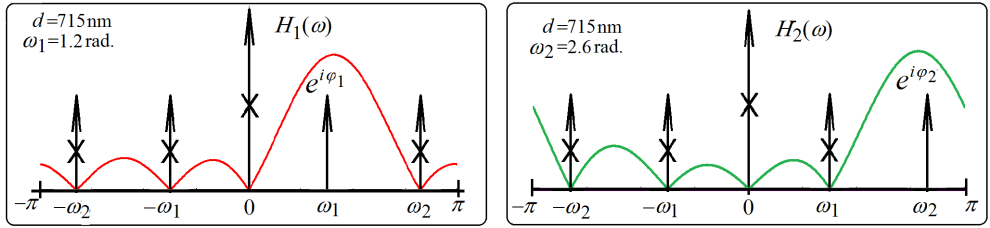

Fig. 8. Spectral plots for $H_1(\omega)$ and $H_2(\omega)$ for the *S/N*-optimized DW-PSA. Note that $\omega_1 = W[(2\pi/\lambda_1)d] = 1.2$ and $\omega_2 = W[(2\pi/\lambda_2)d] = 2.6$; with $W(x) = \arg[\exp(ix)]$.

Therefore in this section we have shown that even though the correct phases $\varphi_1(x, y)$ and $\varphi_2(x, y)$ can be found using Eq. (11) and Eq. (12), without plotting $G_{S/N}(d)$ these DW-PSAs designs will have a low signal-to-noise power-ratio with high probability.

## 9. Example for DW-PSA phase-demodulation for $\lambda_1 = 632.8\,\text{nm}$ and $\lambda_2 = 532.0\,\text{nm}$

Figure 9 shows five computer-simulated interferograms to test the DW-PSAs found in previous section. The PZT-step is $d = 751\,\text{nm}$, giving a good signal-to-noise ratio. As mentioned, for large PZT-steps, the angular frequencies $(\omega_1, \omega_2)$ are wrapped and given by,

$$\omega_1 = \arg\left[\exp(i\,d\,2\pi/\lambda_1)\right] = 2.6, \quad \omega_2 = \arg\left[\exp(i\,d\,2\pi/\lambda_2)\right] = 1.2 \tag{18}$$

Using these angular frequencies in Eq. (11), the specific formula to estimate $\varphi_1(x,y)$ is,

$$A_1(\omega_1)e^{i\varphi_1} = -e^{2.6i}I_0 + (0.78 + 0.62i)I_1 - (0.5 - i)I_2 - (1 + 0.19i)I_3 - e^{-1.4i}I_4 \tag{19}$$

Also, from Eq. (12), the specific 5-step DW-PSA to estimate the signal $\varphi_2(x,y)$ is,

$$A_2(\omega_2)e^{i\varphi_2} = -e^{1.2i}I_0 + (0.8 + 0.6i)I_1 - (0.92 - 0.1i)I_2 + (0.65 - 0.77i)I_3 - e^{1.4i}I_4. \tag{20}$$

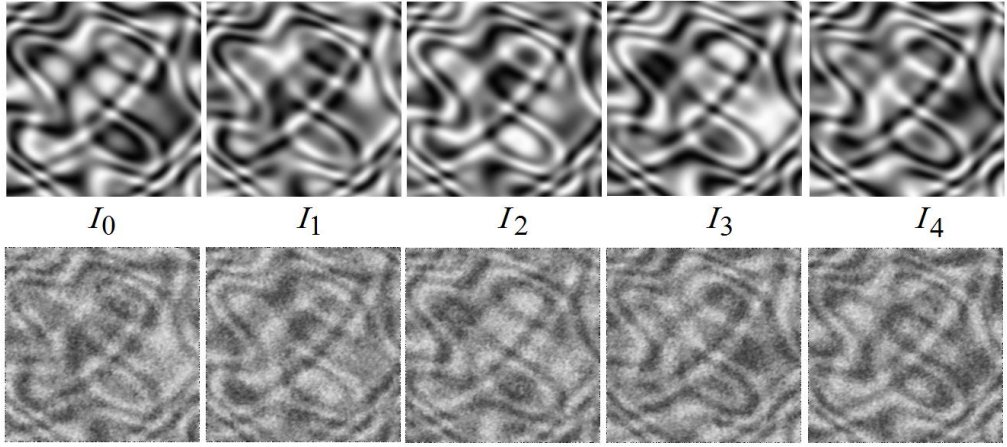

$I_0 \quad I_1 \quad I_2 \quad I_3 \quad I_4$

Fig. 9. The upper row shows 5 simulated interferograms without noise. The lower panel shows the same interferograms corrupted with phase-noise uniformly distributed in $[0,\pi]$. The noisy fringes were low-pass filtered by a 3x3 averaging window.

Figure 10 shows the demodulated signals $\varphi_1(x,y)$ and $\varphi_2(x,y)$.

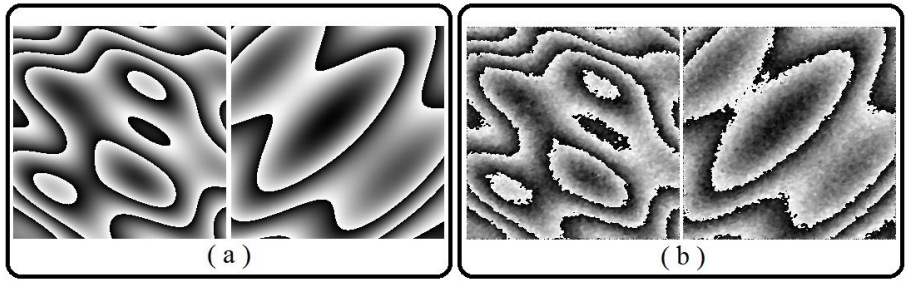

(a) (b)

Fig. 10. The demodulated phases $\varphi_1(x,y)$ and $\varphi_2(x,y)$ corresponding to the noiseless (panel (a)) and noisy (panel (b)) 5-steps interferograms in Fig. 9. Please note that there is no cross-tacking between the two demodulated phases $\varphi_1(x,y)$ and $\varphi_2(x,y)$.

Figure 10(a) shows the noiseless demodulated phases, while Fig. 10(b) shows the demodulated phases degraded with a phase noise uniformly distributed within $[0,\pi]$. Note that absolutely no cross-talking between the demodulated phases $\varphi_1$ and $\varphi_2$ appears.

## 10. Detuning-robust DW-PSA synthesis for $\lambda_1 = 632.8\,\text{nm}$ and $\lambda_2 = 458\,\text{nm}$

Let us assume that our PZT is poorly calibrated. Thus instead of having well-tuned frequencies at $\{\omega_1, \omega_2\}$ we have detuned frequencies at $\{\omega_1 + \Delta, \omega_2 + \Delta\}$, being $\Delta$ the amount of detuning. As Fig. 11 shows, the estimated phase $\hat{\varphi}_2(x,y)$ is now be given by,

$$A_2 e^{-i\hat{\varphi}_2} = H_2(\omega_1 + \Delta)e^{-i\varphi_1} + H_2(\omega_2 + \Delta)e^{-i\varphi_2} + H_2(\omega_1 - \Delta)e^{i\varphi_1} + H_2(\omega_2 - \Delta)e^{i\varphi_2}. \quad (21)$$

The estimated phase $\hat{\varphi}_2(x,y)$ then have cross-talking from $\{e^{-i\varphi_1}, e^{i\varphi_1}, e^{-i\varphi_2}\}$; conversely $\hat{\varphi}_1(x,y)$ will be distorted by cross-talking from $\{e^{-i\varphi_2}, e^{i\varphi_2}, e^{-i\varphi_1}\}$.

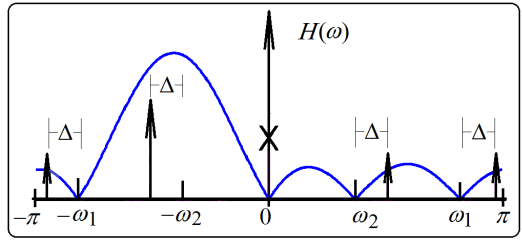

Fig. 11. Here we show the effect of detuning ($\Delta$), greatly exaggerated for clarity. The amount of linear detuning is $\Delta$ (radians/sample). The well-tuned frequencies are $\{-\omega_2, -\omega_2, \omega_1, \omega_2\}$, while the detuned frequencies are $\{-\omega_2 - \Delta, -\omega_2 - \Delta, \omega_1 + \Delta, \omega_2 + \Delta\}$.

To ensure good detuning robustness we need double-zeroes at the rejected frequencies. Therefore, we transform Eq. (8) (5-steps) to detuning-robust DW-FTFs (8-steps) as,

$$H_1(\omega) = \left(1 - e^{i\omega}\right)\left[1 - e^{i(\omega+\omega_2)}\right]^2 \left[1 - e^{i(\omega-\omega_2)}\right]^2 \left[1 - e^{i(\omega+\omega_1)}\right]^2,$$
$$H_2(\omega) = \left(1 - e^{i\omega}\right)\left[1 - e^{i(\omega-\omega_1)}\right]^2 \left[1 - e^{i(\omega+\omega_1)}\right]^2 \left[1 - e^{i(\omega+\omega_2)}\right]^2. \quad (22)$$

Next, we plot $G_{S/N}(d)$ and look for a high local signal-to-noise maximum; see Fig. 12.

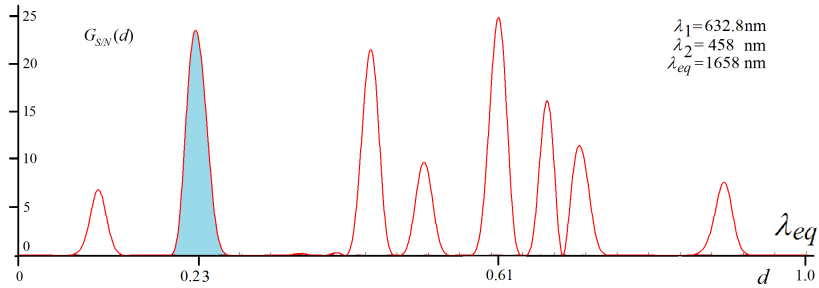

Fig. 12. Graph of the joint signal-to-noise power-ratio $G_{S/N}(d)$ of the two detuning-robust FTF-filters in Eq. (22). The second maximum has a PZT-displacement of $d=381\,\text{nm}$.

We choose the second maximum (in blue) where $G_{S/N}(0.23\lambda_{eq}) = 44$, and $d = 381\,\text{nm}$. Each (8-steps) DW-PSA filter in Eq. (22) has a signal-to-noise ratio of about $\sqrt{44} = 6.6$. Figure 13 shows the two 8-step DW-PSA detuning-robust FTFs. The spectral second-order zeroes are flatter, so they are frequency detuning $\Delta$ tolerant.

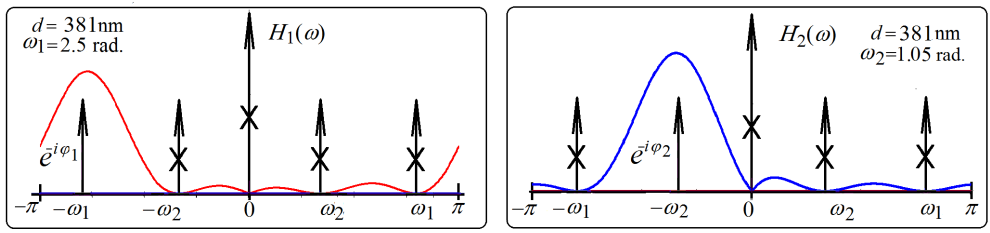

Fig. 13. Spectra of detuning-robust DW-PSA tuned at $\omega_1=2.5\text{rad}$ and $\omega_2=1.05\text{rad}$. The second-order zeroes tolerate a fair amount of frequency detuning $\Delta$.

**11. Harmonic rejection in DW-PSAs**

Figure 14 shows the harmonic response for the FTFs in Eq. (8). The red-sticks are the fringe harmonics at $(n\omega_1)$, and the green ones are the fringe harmonics at $(n\omega_2)$, $|n|\geq 2$.

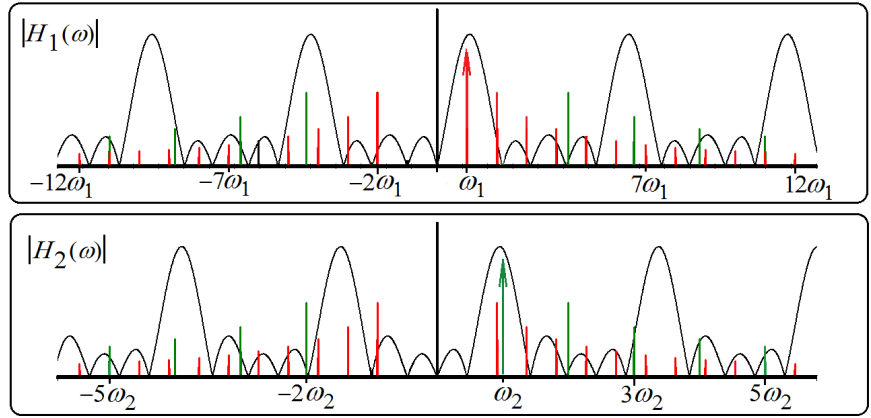

Fig. 14. Amplitudes of the distorting harmonics for $|H_1(n\omega_1)|$, in red; and $|H_2(n\omega_2)|$, in green. The ideal result would be to bandpass just the Dirac's deltas at $\omega=\omega_1$ and $\omega=\omega_2$.

The power of the desired analytic signals $|H_1(\omega_1)\exp(\varphi_1)|^2$ and $|H_2(\omega_2)\exp(\varphi_2)|^2$ with respect to the distorting harmonics is given by,

$$HR_1 = \frac{|H_1(\omega_1)|^2}{\sum_{|n|\geq 2}\left\{\left(\frac{1}{n^2}\right)^2\left[|H_1(n\omega_1)|^2+|H_2(n\omega_2)|^2\right]\right\}} = 11.83,$$

$$HR_2 = \frac{|H_2(\omega_2)|^2}{\sum_{|n|\geq 2}\left\{\left(\frac{1}{n^2}\right)^2\left[|H_1(n\omega_1)|^2+|H_2(n\omega_2)|^2\right]\right\}} = 12.2 \qquad (23)$$

Here we assumed that the amplitude of the harmonics decreases as $(1/n^2)$, so their power decreases as $(1/n^2)^2$. With this assumption, $H_1(\omega_1)$ and $H_2(\omega_2)$ have about 10-times more power than the total power sum of their harmonics $\{H_1(n\omega_1), H_1(n\omega_2), H_2(n\omega_1), H_2(n\omega_2)\}$.

Figure 15 shows five saturated phase-shifted interferograms. These 5 temporal interferograms are then phase demodulated using DW-PSAs, Eqs (11)-(12).

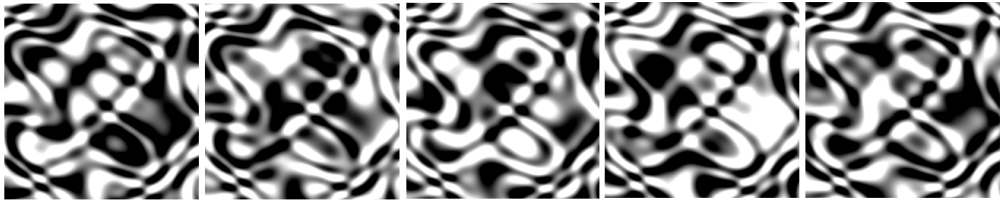

Fig. 15. Five DW phase-shifted temporal interferograms with amplitude saturation.

Figure 16 shows the demodulated phases $\varphi_1$ and $\varphi_2$ of the interferograms in Fig. 15.

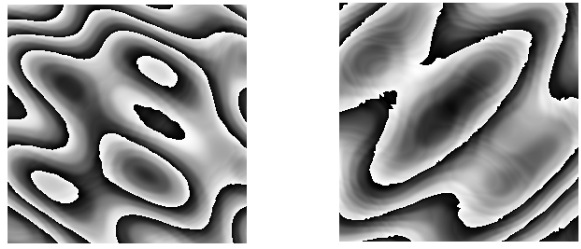

Fig. 16. The two demodulated phases from the 5 saturated fringe patterns in Fig. 15.

## 12. Multi-wavelength $\{\lambda_1, \lambda_2, \lambda_3, ..., \lambda_n\}$ temporal phase-shifting interferometry

Here DW-PSA is generalized to 3-walengths. A simplified schematic of an interferometer simultaneously illuminated with 3-wavelengths $\{\lambda_1, \lambda_2, \lambda_3\}$ is shown in Fig. 17.

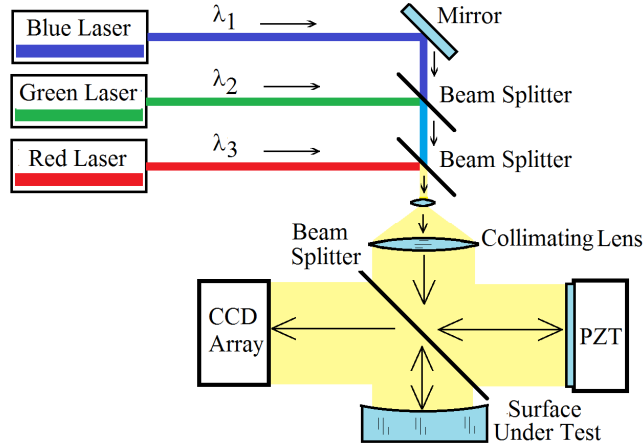

Fig. 17. Simplified schematics for a temporal 3-wavelenght phase-shifting interferometer.

The continuous-time phase-shifted interferogram is,

$$I(x,y,t) = a + b_1 \cos[\varphi_1 + \omega_1 t] + b_2 \cos[\varphi_2 + \omega_2 t] + b_3 \cos[\varphi_3 + \omega_3 t]. \qquad (24)$$

Now we have 7 unknowns $\{a, b_1, b_2, b_3, \varphi_1, \varphi_2, \varphi_3\}$; being $\{\varphi_1, \varphi_2, \varphi_3\}$ the searched phases. Thus we need at least 7 temporal phase-shifted interferograms. Figure 18 shows the Fourier spectrum of this 3-wavelengths interferogram.

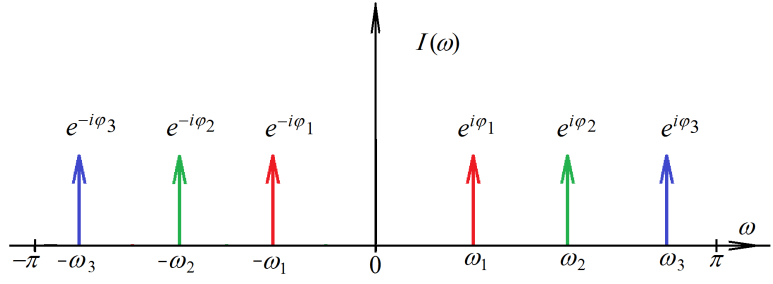

Fig. 18. Fourier spectrum $I(\omega)$ for a 3-wavelength temporal phase-shifted interferogram.

Therefore we need to construct 3-FTFs having at least 6 first-order zeroes (7-steps) as,

$$H_1(\omega) = (1-e^{i\omega})\left[1-e^{i(\omega+\omega_2)}\right]\left[1-e^{i(\omega-\omega_2)}\right]\left[1-e^{i(\omega+\omega_3)}\right]\left[1-e^{i(\omega-\omega_3)}\right]\left[1-e^{i(\omega+\omega_1)}\right],$$
$$H_2(\omega) = (1-e^{i\omega})\left[1-e^{i(\omega-\omega_1)}\right]\left[1-e^{i(\omega+\omega_1)}\right]\left[1-e^{i(\omega+\omega_3)}\right]\left[1-e^{i(\omega-\omega_3)}\right]\left[1-e^{i(\omega+\omega_2)}\right], \quad (25)$$
$$H_3(\omega) = (1-e^{i\omega})\left[1-e^{i(\omega-\omega_1)}\right]\left[1-e^{i(\omega+\omega_1)}\right]\left[1-e^{i(\omega+\omega_2)}\right]\left[1-e^{i(\omega-\omega_2)}\right]\left[1-e^{i(\omega+\omega_3)}\right].$$

The spectrum $H_1(\omega)$ rejects the analytic signals at $\{-\omega_3, -\omega_2, -\omega_1, 0, \omega_2, \omega_3\}$; $H_2(\omega)$ rejects the signals at $\{-\omega_3, -\omega_2, -\omega_1, 0, \omega_1, \omega_3\}$; and $H_3(\omega)$ rejects the analytic signals at $\{-\omega_3, -\omega_2, -\omega_1, 0, \omega_1, \omega_2\}$. Therefore $I(\omega)H_1(\omega)$ passband $\exp(i\varphi_1)\delta(\omega-\omega_1)$; $I(\omega)H_2(\omega)$ passband $\exp(i\varphi_2)\delta(\omega-\omega_2)$, and finally $I(\omega)H_3(\omega)$ passband $\exp(i\varphi_3)\delta(\omega-\omega_3)$.

The joint signal-to-noise power-ratio optimizing criterion now reads,

$$G_{S/N}(d) = \left(\frac{|H_1(\omega_1)|^2}{\frac{1}{2\pi}\int_{-\pi}^{\pi}|H_1(\omega)|^2 d\omega}\right)\left(\frac{|H_2(\omega_2)|^2}{\frac{1}{2\pi}\int_{-\pi}^{\pi}|H_2(\omega)|^2 d\omega}\right)\left(\frac{|H_3(\omega_3)|^2}{\frac{1}{2\pi}\int_{-\pi}^{\pi}|H_3(\omega)|^2 d\omega}\right). \quad (26)$$

We then find a convenient local maximum for $G_{S/N}(d)$, obtaining a fixed PZT-step $d$, and three angular-frequencies $(\omega_1, \omega_2, \omega_3) \in [-\pi, \pi] \times [-\pi, \pi] \times [-\pi, \pi]$ as,

$$\omega_1 = W\left(\frac{2\pi}{\lambda_1}d\right), \quad \omega_2 = W\left(\frac{2\pi}{\lambda_2}d\right), \quad \omega_3 = W\left(\frac{2\pi}{\lambda_3}d\right); \quad W(x) = \arg[\exp(ix)]. \quad (27)$$

The three impulse responses $\{h1(t), h2(t), h3(t)\}$ are then given by,

$$h_1(t) = F^{-1}\{H_1(\omega)\} = \sum_{n=0}^{6} c1_n(\omega_1, \omega_2, \omega_3)\,\delta(t-n),$$
$$h_2(t) = F^{-1}\{H_2(\omega)\} = \sum_{n=0}^{6} c2_n(\omega_1, \omega_2, \omega_3)\,\delta(t-n), \quad (28)$$
$$h_3(t) = F^{-1}\{H_3(\omega)\} = \sum_{n=0}^{6} c3_n(\omega_1, \omega_2, \omega_3)\,\delta(t-n),$$

Here $c1_n(\omega_1, \omega_2, \omega_3)$, $c2_n(\omega_1, \omega_2, \omega_3)$, $c3_n(\omega_1, \omega_2, \omega_3)$ are the complex coefficients of the PSAs, which now depend on the three temporal-carrier frequencies $\{\omega_1, \omega_2, \omega_3\}$.

We now digitally grab 7 phase-shifted interferograms given by:

$$I_n = a + b_1 \cos[\varphi_1 + n\omega_1] + b_2 \cos[\varphi_2 + n\omega_2] + b_3 \cos[\varphi_3 + n\omega_3]; \quad n = 0,\ldots,6. \quad (29)$$

Obtaining the three searched quadrature analytic signals as,

$$A_1(x,y)\exp[i\varphi_1(x,y)] = \sum_{n=0}^{6} c1_n(\omega_1,\omega_2,\omega_3)\, I_n(x,y),$$

$$A_2(x,y)\exp[i\varphi_2(x,y)] = \sum_{n=0}^{6} c2_n(\omega_1,\omega_2,\omega_3)\, I_n(x,y), \quad (30)$$

$$A_3(x,y)\exp[i\varphi_2(x,y)] = \sum_{n=0}^{6} c3_n(\omega_1,\omega_2,\omega_3)\, I_n(x,y),$$

where $A_n(x,y) = (1/2)H_n(\omega_n)b(x,y)$. By mathematical induction, one may see that a 4-wavelength $\{\lambda_1,\lambda_2,\lambda_3,\lambda_4\}$ phase-shifting algorithm would need at least 9 phase-shifted interferograms, requiring FTFs having 8 first–order zeroes, etc, etc.

## 13. Conclusions

The problem that was solved here may be stated as follows: Having a laser interferometer simultaneously illuminated with fixed wavelengths $\{\lambda_1,\lambda_2,...,\lambda_K\}$ and a single PZT phase-shifter, find $K$ phase-shifting algorithms (PSAs) which phase-demodulate $\{\varphi_1,\varphi_2,...,\varphi_K\}$ for each laser-color, with high signal-to-noise and no cross-taking among these phases.

This was solved as follows (for $K$=2, and $K$=3 in section 12),

a) To start, we synthesize two 5-step quadrature-filters (PSA-spectra, Eq. (8)) that bandpass $\exp(i\varphi_1)$ and $\exp(i\varphi_1)$ from 5 phase-shifted interferograms (Eq. (6)) as,

$$H_1(\omega) = \left(1-e^{i\omega}\right)\left[1-e^{i(\omega+\omega_2)}\right]\left[1-e^{i(\omega-\omega_2)}\right]\left[1-e^{i(\omega+\omega_1)}\right],$$
$$H_2(\omega) = \left(1-e^{i\omega}\right)\left[1-e^{i(\omega-\omega_1)}\right]\left[1-e^{i(\omega+\omega_1)}\right]\left[1-e^{i(\omega+\omega_2)}\right]. \quad (31)$$

b) We then jointly optimize $\{H_1(\omega),H_2(\omega)\}$ for high signal-to-noise ratio $G_{S/N}(d)$ (Fig. 7) and obtained the PZT-step $d$ at which that local maximum occurs.

c) Having an optimum PZT-step $d$, we then calculate the tuning frequencies $\omega_1 = (2\pi/\lambda_1)d$, $\omega_2 = (2\pi/\lambda_2)d$, which substituted back into $\{H_1(\omega),H_2(\omega)\}$ give the specific DW-PSAs that phase-demodulate $\varphi_1(x,y)$ and $\varphi_2(x,y)$ (Eqs. (11)-12)).

d) We then plot (Fig. 8) the *S/N*-optimized designs $\{H_1(\omega),H_2(\omega)\}$ to gauge their spectral behavior within $\omega \in [-\pi,\pi]$ (Fig. 8).

e) We also plotted (Fig. 14) the *S/N*-optimized $\{H_1(\omega),H_2(\omega)\}$ designs for an extended frequency range of $\omega \in [-30\pi,30\pi]$, to gauge their harmonic-rejection.

f) With the *S/N*-optimized $\{H_1(\omega),H_2(\omega)\}$ designs we quantified the harmonic-rejection capacity for each DW-PSA (Eq. (23)).

g) For poor PZT-calibration we modified $\{H_1(\omega),H_2(\omega)\}$ by raising the first-order zeroes to second-order zeroes, i.e. $(\omega-\omega_1) \Rightarrow (\omega-\omega_1)^2$, $(\omega-\omega_2) \Rightarrow (\omega-\omega_2)^2$, etc.; making $\{H1(\omega),H2(\omega)\}$ robust to detuning at the rejected frequencies (Fig. 13).

h) We used the *S/N*-optimized designs to phase-demodulate 5 phase-shifted interferograms (Figs. 9-10) with high signal-to-noise and no phase cross-talking.

i) Finally we extended the DW-PSA theory to 3-wavelengths $\{\lambda_1,\lambda_2,\lambda_3\}$; further $K$-wavelengths $\{\lambda_1,\lambda_2,...,\lambda_K\}$ generalization of this theory is just a matter of mathematical induction.

Finally, two examples of DW-PSA demodulation with $\{\lambda_1 = 632.8\text{nm}, \lambda_2 = 532\text{nm}\}$ which illustrate the behavior of the synthesized PSAs were given. As far as we know, previous art on DW-PSAs [2-5,14-19] only provided *ad-hoc* multi-wavelength PSA designs. Thus, this is the first time that a general theory for synthesizing and analyzing multi-wavelength temporal phase-shifting algorithms is presented, and from which one may derive quantifying formulas for: (a) the PSAs spectra for each wavelength, (b) the PSAs signal-to-noise robustness for each wavelength, (c) the PSAs detuning sensitivity, and (d) the PSAs harmonics rejection for each wavelength.

**Acknowledgments**

The authors acknowledge the financial support of the Mexican National Council for Science and Technology (CONACYT), grant 157044. Also the authors acknowledge Cornell University for supporting the e-print repository arXiv.org and the Optical Society of America for permitting OSA's contributors to post their manuscript at arXiv.